\documentclass[nobm]{epl}

\usepackage{graphicx}
\usepackage{amssymb}
\usepackage{latexsym}

\newcommand{\fig}[1]{Fig.~\ref{#1}}

\newcommand{\bd}{{\rm b}}
\newcommand{\piad}{\pi_{\rm ad}}

\def\siml{\mathrel{\mathchoice {\vcenter{\offinterlineskip\halign{\hfil
$\displaystyle##$\hfil\cr<\cr\sim\cr}}}
{\vcenter{\offinterlineskip\halign{\hfil$\textstyle##$\hfil\cr
<\cr\sim\cr}}}
{\vcenter{\offinterlineskip\halign{\hfil$\scriptstyle##$\hfil\cr
<\cr\sim\cr}}}
{\vcenter{\offinterlineskip\halign{\hfil$\scriptscriptstyle##$\hfil\cr
<\cr\sim\cr}}}}}
\def\simg{\mathrel{\mathchoice {\vcenter{\offinterlineskip\halign{\hfil
$\displaystyle##$\hfil\cr>\cr\sim\cr}}}
{\vcenter{\offinterlineskip\halign{\hfil$\textstyle##$\hfil\cr
>\cr\sim\cr}}}
{\vcenter{\offinterlineskip\halign{\hfil$\scriptstyle##$\hfil\cr
>\cr\sim\cr}}}
{\vcenter{\offinterlineskip\halign{\hfil$\scriptscriptstyle##$\hfil\cr
>\cr\sim\cr}}}}}

\title{Phase transitions in systems with two species of molecular
  motors}
\shorttitle{Two species of molecular motors}
\author{Stefan Klumpp\and Reinhard Lipowsky}
\institute{Max-Planck-Institut f\"ur Kolloid- und
  Grenzfl\"achenforschung -- 14424 Potsdam, Germany
}
\pacs{64.60.Cn}{Order-disorder transformations; statistical mechanics of model systems}
\pacs{82.70.-y}{Disperse systems; complex fluids}
\pacs{87.16.Uv}{Active transport processes}

\begin{document}

\maketitle

\begin{abstract}
  Systems with two species of active molecular motors moving on
  (cytoskeletal) filaments into opposite directions are studied
  theoretically using driven lattice gas models. The motors can unbind
  from and rebind to the filaments. Two motors are more likely to bind
  on adjacent filament sites if they belong to the same species. These
  systems exhibit (i) Continuous phase transitions towards states with
  spontaneously broken symmetry, where one motor species is largely
  excluded from the filament, (ii) Hysteresis of the total current
  upon varying the relative concentrations of the two motor species,
  and (iii) Coexistence of traffic lanes with opposite directionality
  in multi-filament systems. These theoretical predictions should be
  experimentally accessible.
\end{abstract}


Cytoskeletal motors which convert chemical free energy into directed
movements along filament tracks have been studied extensively during
the last decade. Many different motor molecules have been identified
and much insight has been gained into their motor mechanisms and their
functions within cells \cite{Schliwa_Woehlke2003}. Much of the present
knowledge about motor mechanisms has been obtained from \emph{single}
molecule experiments, which have been performed under a variety of
different conditions \cite{Howard2001}.

\emph{Cooperative} effects, on the other hand, arising from
motor--motor interactions have hardly been explored experimentally so
far.  Most of the available information about these interactions is
due to decoration experiments where the binding patterns of inactive
(non-moving) motors to immobilized filaments are examined by electron
microscopy and X-ray scattering.
Decoration experiments clearly demonstrate mutual exclusion from
binding sites of the filaments. In addition, there is evidence for an
effectively attractive motor--motor interaction mediated via the
filament. Such an interaction is implied by the coexistence of
decorated and bare filaments, which has been observed both for the
decoration of actin filaments by myosin
\cite{Woodrum__Pollard1975,Orlova_Egelman1997} and for the decoration
of microtubules by kinesin \cite{Vilfan__Mandelkow2001}.  In the case
of actin decoration, the motor--motor interaction depends on the
internal conformation of the actin filaments
\cite{Orlova_Egelman1997}.  This observation as well as experimental
results on active kinesin in the presence of ATP \cite{Muto2001}
suggest that a bound motor leads to a localized deformation of the
filament which promotes the binding of further motors on adjacent
binding sites.

A convenient way to study interacting motors theoretically is to model
the motor movements as walks on a lattice
\cite{Lipowsky__Nieuwenhuizen2001,Klumpp_Lipowsky2003}. In these
models, the directed walks along filaments are described as biased
random walks whereas the unbound motors undergo symmetric random walks
corresponding to diffusive motion. Interactions can be taken into
account by hopping rates which depend on the presence of motors on
adjacent lattice sites.  These models are new variants of driven
lattice gas models or exclusion processes, where the driving is
localized to the filaments.  Other variants of driven lattice gas
models have been previously studied for a variety of transport
processes, see, e.g.\ \cite{Krug1991,Schuetz2001,Evans__Mukamel1995}.

In this Letter, we will address the cooperative behavior of two
species of motors moving along the same filament, but into opposite
directions. In both the kinesin and myosin motor families, motors
moving into opposite direction have been identified, e.g.\ myosin V
and VI or conventional kinesin and ncd \cite{Schliwa_Woehlke2003}. The
directionality of motors can also be engineered by genetic methods,
e.g.\ \cite{Henningsen_Schliwa1997,Sablin__Fletterick1998}.  The two
species of motors compete for the same binding sites along the
filament \cite{Lockhard__Cross1995}.  In order to incorporate the
effectively attractive motor--motor interactions as deduced from the
decoration experiments
\cite{Woodrum__Pollard1975,Orlova_Egelman1997,Vilfan__Mandelkow2001,Muto2001},
a bound motor is taken to increase the adsorption rate onto adjacent
binding sites for motors of the same species.  We will show that these
models exhibit states of spontaneously broken symmetry, hysteresis of
current and motor densities, and, in multi-filament systems, the
coexistence of traffic lanes with opposite directionality.  In
contrast to the non-equilibrium phase transitions of the asymmetric
simple exclusion process \cite{Krug1991} or the 'bridge' model
\cite{Evans__Mukamel1995},
the transitions discussed here are \emph{not} induced by the
boundaries, but by the binding and unbinding dynamics of the active
particles. This implies that they can be simply controlled by the bulk
concentrations of the two motor species.

To proceed, let us consider a one-dimensional lattice representing a
filament on which two species of motor particles move. We will denote
the two species by 'plus' and 'minus'. Plus-motors attempt to hop to
the right with a rate $\alpha=v_\bd/\ell$, which depends on the motor
velocity $v_\bd$ in the absence of other motors and on the filament
repeat distance $\ell$, while minus-motors move to the left with the
same rate $\alpha$.  Backward steps are neglected, because they are
rare for cytoskeletal motors. Hopping attempts are only successful if
the target site is not occupied by another motor.  This incorporates
the mutual exclusion of the motors and prevents, in particular, that
bound plus- and minus-motors pass each other.

In general, the filament is located within a larger volume such as a
tube with a certain amount of unbound motors which can bind to an
empty filament site. Likewise, bound motors have a small probability
to unbind at each step. In order to model this system, we replace the
solution surrounding the filament by a reservoir which is
characterized by the two concentrations $\rho_{+}$ and $\rho_{-}$ of
the unbound plus- and minus-motors. Very similar behavior is found for
a refined model as in \cite{Klumpp_Lipowsky2003} where the unbound
motors undergo symmetric random walks within a tube and the number of
particles is locally conserved.

The rates for binding and unbinding depend on the state of adjacent
lattice sites. If no other motors are close, an unbound plus- or
minus-motor binds to an empty filament site with rate $\piad
\rho_{\pm}$, where $\piad$ is the adsorption rate and $\rho_{\pm}$ are
the concentrations of unbound plus- and minus-motors in the solution,
and a bound motor unbinds with rate $\epsilon$.\footnote{As in
  {\cite{Lipowsky__Nieuwenhuizen2001,Klumpp_Lipowsky2003}}, we do
  \emph{not} scale the detachment/attachment rates with the system
  size $L$, in contrast to \cite{fussn1}.} A typical value for
$\epsilon$ is $\alpha/100$ as follows from the observed walking
distances \cite{Lipowsky__Nieuwenhuizen2001}.\footnote{For motor
  complexes or cargo particles with $n$ active motors, one would have
  $\epsilon\sim\epsilon_1^n$ with the detachment rate $\epsilon_1$ of
  a single motor. In addition, the velocity, and thus $\alpha$, are
  reduced for such a particle due to the mutual hindrance of the
  active motors.}

\begin{figure}[tb]
  \onefigure[angle=0,width=.8\textwidth]{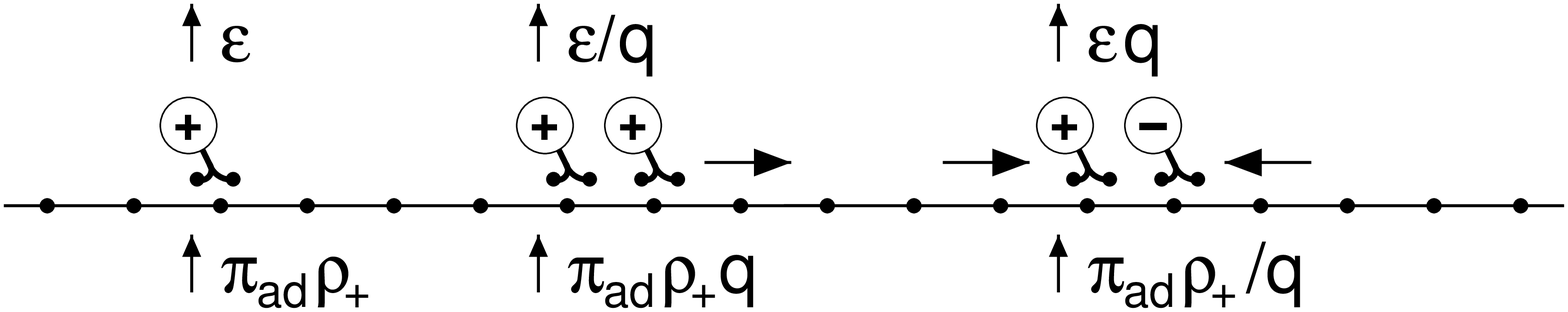}
     \caption{'Plus' and 'minus' motors which  
       move on the filament to the right and to the left,
       respectively.  The rates for attachment to and detachment from
       the filament depend on the state of the forward neighbor site
       on the filament.  (Left) At sites with a vacant forward
       neighbor, motors unbind and bind with rates $\epsilon$ and
       $\piad \rho_{\pm}$; (Middle) A motor of the same species at the
       forward neighbor site enhances binding and reduces unbinding by
       factors $q>1$ and $1/q$, respectively; (Right) A motor of the
       other species reduces binding and enhances unbinding.}
     \label{fig:ww2_cartoon}
\end{figure}

If the motors interact {\em only} via their mutual exclusion, {\em no}
phase transition occurs in the system as one varies the bulk motor
concentrations.  Such transitions are found, however, if one
incorporates the previously mentioned filament-mediated interaction
which will affect, in general, both the binding and the unbinding
rates, $\pi_{ad}$ and $\epsilon$, in the direction perpendicular to
the filament and the forward rate $\alpha$ for steps parallel to the
filament.

Let us assume that the binding rate $\pi_{ad}$ is increased by a
factor $q$ and that the unbinding rate $\epsilon$ is decreased by a
factor $1/q$ if another motor of the same species already occupies the
forward neighbor site on the filament, see
\fig{fig:ww2_cartoon}.\footnote{Interactions involving nearest
  neighbors in both
  forward and backward direction lead to similar results.}
These binding and unbinding processes obey detailed balance
\cite{Klumpp_Lipowsky2003}.  For steps along the filament, on the
other hand, detailed balance is broken since these active steps are
coupled to ATP hydrolysis, and the corresponding rate $\alpha$ will,
in general, change to $\alpha/q^\prime$ with $q^\prime \ne q$ if a
motor of the same species is present on the forward neighbor of the
target site.  We find that the system undergoes a phase transition for
fixed $q^\prime$ and sufficiently large values of $q$. In order to
eliminate one parameter, we will focus in the following on the
situation with $q^\prime = 1$.  For an effectively attractive
interaction between two motors of the same species, we have $q>1$.  In
the presence of a bound motor of the other species at the forward
neighbor site the unbinding rate is enhanced by a factor $q$, while
the adsorption rate is reduced by a factor $1/q$.

In order to suppress all effects arising from the two ends of the
filament, we will first consider periodic boundary conditions in the
longitudinal direction parallel to the filament. Therefore the phase
transitions found here can \emph{not} be induced by the boundary
conditions imposed on the system in contrast to
\cite{Krug1991,Evans__Mukamel1995}.

\begin{figure}[tb]
  \onefigure[angle=-90,width=.8\textwidth]{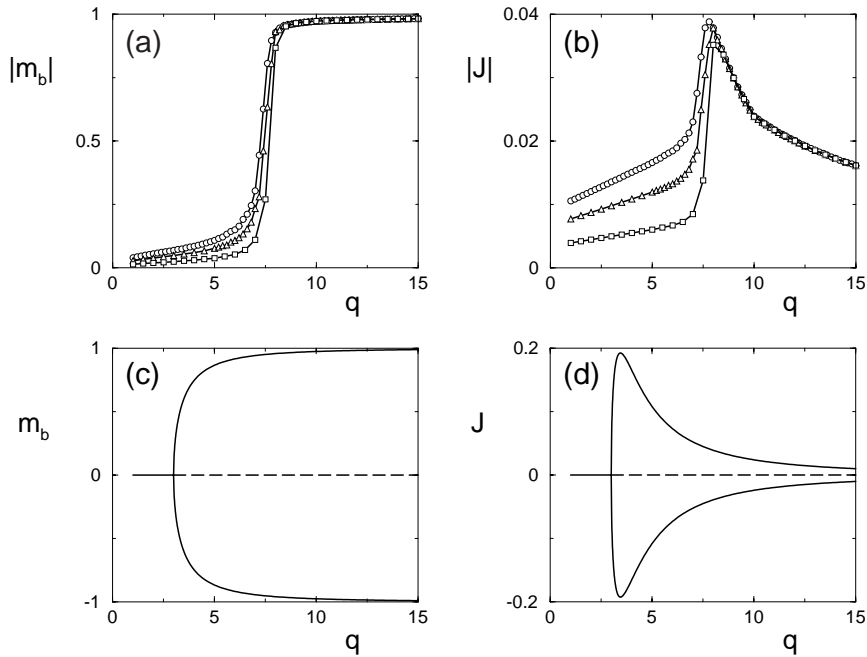}
     \caption{Density difference 
       $m_\bd=\rho_{\rm b,+}-\rho_{\rm b,-}$ and total current
       $J=J_{+}+J_{-}$ as functions of the interaction parameter $q$
       as obtained from MC simulations (a,b) and analytical mean field
       calculations (c,d). In (a,b) the absolute values of $m_\bd$ and
       $J$ are shown for different filament lengths $L=200$ ($\circ$),
       $L=400$ ($\triangle$), and $L=1600$ ($\Box$), see text. The
       mean field solutions (c,d) show a symmetric solution $m_\bd=0$
       and $J=0$ and two branches with broken symmetry for $q>q_c$.
       The hopping rates are $\alpha=1$, $\epsilon=0.01$, $\piad=0.1$,
       and the unbound motor concentrations are
       $\rho=2\rho_{+}=2\rho_{-}=0.1$. }
     \label{fig:2sortenSIM_mb_JvonF}
\end{figure}

For large $q$, motors bound to the filament strongly attract other
motors of the same type to the filament, but repel those of the other
type. Now, if the bound concentration of one motor species is much
higher than the one of the other species, the second species is
basically excluded from the binding sites of the filament. On the
other hand, the motors of the majority species on the filament are
unlikely to unbind, since they are effectively attracted by their
neighbors.  In fact, these interactions lead to a
\emph{non-equilibrium phase transition} induced by the active
movements of the motors with a critical point at $q=q_c$ for
$\rho_{+}=\rho_{-}=\rho/2$, i.e., for equal bulk concentrations of
plus- and minus-motors. We have found these phase transitions both
from Monte Carlo (MC) simulations and analytical mean-field
calculations, compare \fig{fig:2sortenSIM_mb_JvonF}.

For $\rho_+=\rho_-=\rho/2$ and $q<q_c$, motors of both species are
bound to the filament and the densities of bound motors, $\rho_{\rm
  b,+}$ and $\rho_{\rm b,-}$, are equal for both types. The total
current vanishes as the motor currents with opposite directionality,
$J_{+}$ and $J_{-}$, balance each other.  For $q>q_c$, on the other
hand, we observe spontaneous symmetry breaking. Motors of one species
are bound to the filament, while those of the other species are
largely excluded from it, resulting in a non-zero value of the density
difference $m_\bd\equiv\rho_{\rm b,+}-\rho_{\rm b,-}$ as shown in
\fig{fig:2sortenSIM_mb_JvonF}(a) and (c). Likewise the total current
$J\equiv J_{+}+J_{-}$ is also finite for $q>q_c$.  The current
decreases for large $q$ as the filament becomes overcrowded. The MC
simulations exhibit strong finite-size effects. We sometimes observe
reversal of the direction of the current at $q>q_c$, as the system
flips from one broken symmetry state to the other.  In addition, for
values of $q$ close to $q_c$, the total current $J$ and the difference
$m_\bd$ of bound densities exhibit strong fluctuations.  Because of
these fluctuations, we plotted the absolute values of $m_\bd$ and $J$
in \fig{fig:2sortenSIM_mb_JvonF}(a) and (b). These quantities exhibit
non-zero values for finite system size $L$, but decrease to zero for
large $L$.

\begin{figure}[tb]
  \onefigure[angle=-90,width=.8\textwidth]{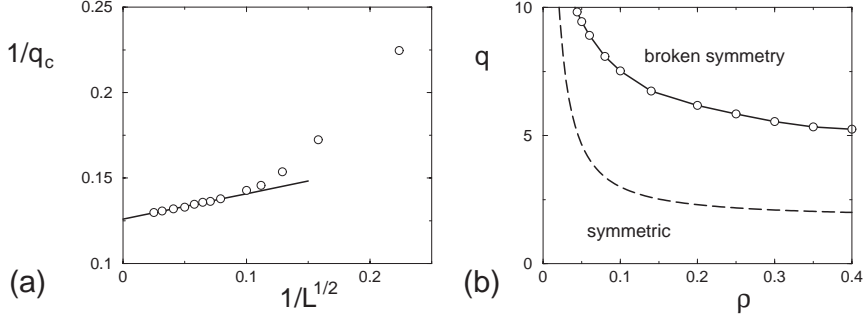}
     \caption{(a) The inverse critical interaction strength $1/q_c$ 
       as obtained from simulations for different system sizes $L$
       with $20\leq L\leq 1600$.  The line is a linear fit to the data
       which leads to $q_c=7.9\pm 0.1$ at infinite $L$. (b) Phase
       diagram as a function of the interaction parameter $q$ and the
       concentration of unbound motors $\rho=2\rho_{\pm}$, as obtained
       from simulations with $L=400$ (data points) and from mean field
       theory (dashed line).  Parameters are as in
       \fig{fig:2sortenSIM_mb_JvonF}.  }
     \label{fig:phasendiagramm}
\end{figure}

The critical interaction parameter $q_c$ as determined by simulations
depends on $L$, see \fig{fig:phasendiagramm}(a). Extrapolation of
these data leads to the estimate $q_c= 7.9\pm 0.1$ for infinite $L$
and to $q_c(\infty)-q_c(L)\sim 1/L^{1/2}$.  In addition, we find that
the time between subsequent flips from one ordered state to the other
at $q>q_c$ increases exponentially with $L$ \cite{Klumpp_LipowskyTBP}.
These observations strongly indicate that symmetry breaking persists
in the thermodynamic limit.

We have determined the critical interaction parameter $q_c$ as a
function of the overall motor concentration $\rho\equiv
\rho_{+}+\rho_{-}$ in the solution. The resulting phase diagram as
obtained from MC simulations and mean-field calculations is shown in
\fig{fig:phasendiagramm}(b).

In systems with several parallel filaments, the symmetry breaking
leads to the coexistence of traffic lanes with opposite
directionality. We have performed simulations for the case of two
parallel filaments which are placed within a cylindrical tube,
parallel to the cylinder axis \cite{Klumpp_LipowskyTBP}. In this case,
the particle number is locally conserved.  Therefore, if one motor
species starts to decorate one filament, the other motor species
attains a larger bulk concentration and is, thus, more likely to bind
to the other filament. Indeed, for $q>q_c$, we observe that the two
filaments are covered by different motor species which then form two
traffic lanes with opposite directionality. Thus, the symmetry
breaking provides a simple mechanism for efficient transport between
two reservoirs of cargo particles.

Within mean-field theory, the time evolution of the densities of bound
motors $\rho_{\bd,+}$ and $\rho_{\bd,-}$ is given by
\begin{equation}\label{eq:timeEvolution}
  \frac{\partial}{\partial t}  \rho_{\rm b,\pm}  +\frac{\partial}{\partial x}J_{\pm} =I_{\pm}(q),
\end{equation}
with the currents
\begin{equation}
  J_{\pm}\equiv\pm v_\bd \rho_{\rm b,\pm}(1-\rho_\bd)- D_\bd \left(\frac{\partial\rho_{\rm b,\pm}}{\partial x} + \rho_{\rm b,\pm}\frac{\partial \rho_{\rm b,\mp}}{\partial x} -\rho_{\rm b,\mp}\frac{\partial \rho_{\rm b,\pm}}{\partial x} \right)
\end{equation} 
and the attachment/detachment terms
\begin{eqnarray}
  \lefteqn{I_{\pm}(q)  \equiv  {}-\epsilon\rho_{\rm b,\pm}(1-\rho)\Big[(1-\rho_\bd)+ q\rho_{\rm b,\mp}+\frac{1}{q}\rho_{\rm b,\pm}\Big] } \nonumber\\
   & & {}+\piad \rho_{\pm}(1-\rho_\bd)\Big[(1-\rho_\bd)+ \frac{1}{q}\rho_{\rm b,\mp}+q\rho_{\rm b,\pm}\Big].\label{eq:BindingUnbinding}
\end{eqnarray}
Here the parameters $v_\bd$ and $D_\bd$ are the velocity and the
diffusion coefficient of single bound motors and
$\rho_\bd\equiv\rho_{\rm b,+}+\rho_{\rm b,-}$.  The factor $(1-\rho)$
in the detachment term describes crowding of motors in the solution
with $(1-\rho)\approx 1$ for typical experimental conditions.  It
follows from (\ref{eq:timeEvolution})--(\ref{eq:BindingUnbinding})
that the stationary and spatially homogeneous states satisfy
$I_{+}=I_{-}=0$.  For the symmetric case with
$\rho_{+}=\rho_{-}=\rho/2$, this leads to a critical point at $q=q_c$
as given by
\begin{equation}
  \label{eq:qc}
  q_c\equiv \eta+\sqrt{\eta^2+3}\qquad{\rm with}\qquad \eta\equiv\frac{\epsilon(1-\rho)}{\piad\rho}.
\end{equation}
The corresponding order parameter $m_\bd\equiv\rho_{\rm b,+}-\rho_{\rm
  b,-}$ vanishes for $q<q_c$, but attains the finite value
\begin{equation}
  m_\bd=\pm \sqrt{\frac{(q^2-2 \eta q-3)(q^2-2\eta q+1)}{(q-1)^2(q-2\eta+1)^2} }\sim\pm(q-q_c)^{1/2}
\end{equation}
for $q>q_c$. Likewise, the total current $J=v_\bd m_\bd(1-\rho_\bd)$
also vanishes as $(q-q_c)^{1/2}$. The analysis of the MC data
indicates that $m_\bd\sim (q-q_c)^{\beta}$ with $\beta\simeq 0.35$.

Linear stability analysis shows that the symmetric solution with
$m_\bd=0$ is stable for $q<q_c$ and becomes unstable for $q>q_c$,
while the solution with broken symmetry is stable in the range of $q$
where it exists.\footnote{Our mean field equations also imply a
  spurious 'blocking transition' for small $q$ and large $v_\bd$ which
  is \emph{not} observed in our MC simulations, compare
  \cite{fussn2}.}

\begin{figure}[tb]
  \onefigure[angle=-90,width=.8\textwidth]{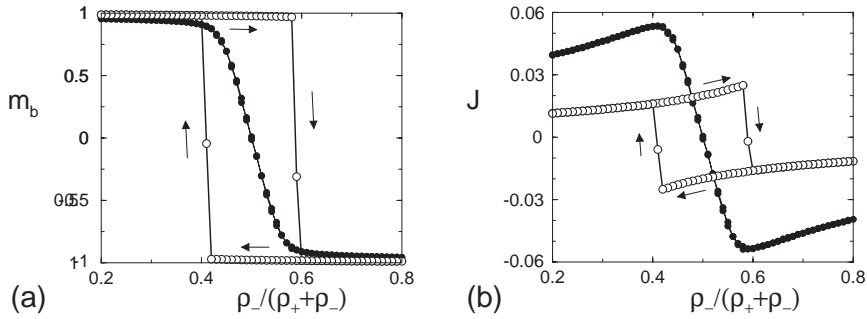}
     \caption{(a) The density difference 
       $m_\bd=\rho_{\rm b,+}-\rho_{\rm b,-}$ and (b) the total current
       $J=J_{+}+J_{-}$ as a function of the fraction of minus-motors
       $\rho_{-}/\rho$ for fixed overall concentration,
       $\rho=\rho_{+}+\rho_{-}=0.1$.  The white and black data points
       correspond to $q=12>q_c\simeq 7.5$ and $q=6<q_c$, respectively.
       For $q<q_c$, no hysteresis is observed.  Data points have been
       obtained by simulating $10^6$ MC steps for each value of
       $\rho_{-}$.  The rates are as in \fig{fig:2sortenSIM_mb_JvonF}
       and the filament length $L=400$.}
     \label{fig:HystereseJ}
\end{figure}

Varying the difference $\rho_{+}-\rho_{-}$ of the unbound motor
concentrations for constant overall concentration
$\rho=\rho_{+}+\rho_{-}$, we observe a discontinuous transition with
hysteresis at $q>q_c$, see \fig{fig:HystereseJ}. It is interesting to
note that the total current, $J=J_{+}+J_{-}$, increases as the
concentration of minority motors, which are excluded from the
filament, is increased, and thus adopts its maximal value in the
region of metastability.  To observe hysteresis, it is, however, not
necessary to keep the overall concentration $\rho$ constant.  In an
experiment, one could start with $\rho_{-}<\rho_{+}$, add first
minus-motors until $\rho_{-}>\rho_{+}$ and then add plus-motors,
thereby increasing the overall concentration $\rho$. Such an
experiment will again show hysteresis since $q_c$ decreases with
increasing $\rho$, as shown in \fig{fig:phasendiagramm}.

The systems discussed so far were characterized by periodic boundary
conditions in the longitudinal direction parallel to the filament.
These boundary conditions are useful from a theoretical point of view
since they suppress all boundary effects which can dominate the
relatively small systems accessible to computer simulations.  In
principle, one could study such systems experimentally if one combined
several filaments to ring-like arrangement. In practice, essentially
straight filaments of finite length with two `open' ends are simpler
to prepare.  Thus, let us finally consider such systems consisting of
one or several straight filaments in contact with a solution of plus-
and minus-motors, again with the concentrations $\rho_{+}$ and
$\rho_{-}$, respectively. Motors can attach to the filament and unbind
from it with the same rates as before with the exception of the last
binding sites at the filament ends, where unbinding occurs with rate
$\epsilon_{\rm end}$.  If the filament is sufficiently long with
length $L\gg v_\bd q/\epsilon$ and $L\gg v_\bd q/\epsilon_{\rm end}$,
the bound density along the filament will be determined by the bulk
dynamics, while the dynamics at the filament ends leads to the
formation of boundary layers, so that the phase transitions described
above for the periodic case also occur for the case with 'open' ends.
In order to see whether these phase transitions could be observed in
experiments where the typical filament length is tens of micrometers,
we performed simulations for this open case with $L=1000$ and motor
parameters as in \fig{fig:2sortenSIM_mb_JvonF}. In these simulations,
symmetry breaking and hysteresis are clearly observed for an
intermediate range of the unbinding rate at the filament end
$\epsilon_{\rm end}$ as given by $0.03\protect\siml\epsilon_{\rm
  end}\protect\siml 0.5$,
with approximately the same value of $q_c$ as for the periodic
case.\footnote{For smaller or larger $\epsilon_{\rm end}$, the
  boundary layers become quite large and the density difference
  $m_\bd$ fluctuates strongly at intermediate $q$. For small
  $\epsilon_{\rm end}$, plus-motors crowd at the plus-end and
  minus-motors at the minus-end.  For large $\epsilon_{\rm end}$,
  plus-motors are depleted at the plus-end, which is then dominated by
  minus-motors, and vice versa at the minus-end.} Since $\epsilon_{\rm
  end}$ can be expected to lie within this range for cytoskeletal
motors, this implies that these phase transitions should be accessible
in {\it in vitro} experiments with two species of motors and filaments
of length $L\protect\simg 10\mu$m.

In summary, we have discussed models for two species of interacting
molecular motors. Binding to a filament is promoted by bound motors of
the same, but suppressed by bound motors of the other species. These
systems exhibit active, non-equilibrium phase transitions with
spontaneous symmetry breaking, hysteresis and (in the case of many
filaments) the formation of lanes with opposite directionality. On the
one hand, these systems represent new lattice gas models, which
exhibit phase transitions between two ordered states induced by active
particles.  On the other hand, these novel phase transitions should
also be accessible to experiments with cytoskeletal motors.

\end{document}